# Realization of bound states in the continuum in anti-*PT*-symmetric optical systems

Jiawei Zhang, Ziyao Feng, and Xiankai Sun[*]

*Department of Electronic Engineering, The Chinese University of Hong Kong, Shatin, New Territories, Hong Kong*

[*]*Corresponding author: xksun@cuhk.edu.hk*

**Novel physical concepts that originate from quantum mechanics, such as non-Hermitian systems (dealing mostly with *PT* and anti-*PT* symmetry) and bound states in the continuum (BICs), have attracted great interest in the optics and photonics community. To date, BICs and anti-*PT* symmetry seem to be two independent topics. Here, we for the first time propose a parallel cascaded-resonator system to achieve BICs and anti-*PT* symmetry simultaneously. We found that the requirements for the Fabry–Pérot BIC and anti-*PT* symmetry can both be satisfied when the phase shift between any two adjacent resonators is an integer multiple of $\pi$. We further analyzed the cascaded-resonator systems which consist of different numbers of resonators and demonstrated their robustness to fabrication imperfections. The proposed structure can readily be realized on an integrated photonic platform, which can have many applications that benefit from the advantages of both BICs and anti-*PT* symmetry, such as ultralow-linewidth lasing, enhanced optical sensing, and optical signal processing.**

## Introduction

Non-Hermitian physics, which exhibits properties significantly different from that of a closed system, has attracted increasing interest[1, 2]. In 1998, *PT* symmetry was proposed by Bender and Boettcher in quantum mechanics to investigate non-Hermitian systems with real eigenvalues[3]. *PT*-symmetric systems are invariant under the combined action of parity reversal (*P*) and time-reversal (*T*) operations. The exceptional point separates the *PT*-symmetric phase (with real eigenvalues) from the *PT*-broken phase (with complex eigenvalues). *PT*-symmetric systems have been investigated and demonstrated in various areas[4–10]. Among them, optical systems can provide ideal platforms for non-Hermitian physics with tunability and accessibility. Optical realizations of *PT*-symmetric systems not only advance the theoretical studies of non-Hermitian physics, but also give rise to various applications in photonics, including *PT*-symmetric lasers[11–13], high-sensitivity sensing at exceptional points[14–16], and nonreciprocal light propagation[17, 18].

Anti-*PT* (*APT*) symmetry, the counterpart of *PT*-symmetry, imposes that the Hamiltonian anticommutes with the joint parity and time operator. Unlike the *PT*-symmetric systems[19], *APT*-symmetric systems can be realized in the absence of a gain medium, which makes such systems much more accessible[20, 21]. To date, *APT* symmetry has been observed and investigated in cold atoms[22, 23], microwaves[24, 25], nonlinear optical systems[26, 27], and integrated photonics[20, 21]. In integrated photonics, there are some easy-to-implement configurations for realizing *APT* symmetry, including optical waveguides and microcavities[20, 21].

Bound states in the continuum initially proposed in quantum mechanics by von Neumann and Wigner refer to a wave state which is perfectly confined without any radiation loss even though it exists in a continuous spectrum[28]. BICs have been investigated and demonstrated in electromagnetic, acoustic, and water waves[29]. Recently, the development of nanofabrication technologies has enabled realizations of BICs in photonics[30–34], leading to many new applications such as enhanced optical nonlinearity[35], lasers[32, 36, 37], filters[38], and sensors[37, 39]. In integrated photonics, BICs can be achieved by coupling resonators to the radiation reservoir with perfectly destructive interference among the dissipation channels[29].

To date, BICs and *APT* symmetry seem to be two independent topics. Here, we propose a parallel cascaded-resonator system to achieve BICs and *APT* symmetry simultaneously. For *APT*-symmetric optical systems consisting of $N$ ($N > 2$) resonators, we first theoretically studied the evolution of eigenfrequencies, as well as the transmission characteristics. Then, we numerically simulated these structures with a finite-element method, and the simulated results agree well with the analytical results. We also compared the cascaded-resonator systems, which consist of different numbers of resonators, and analyzed the effects of experimental imperfections. The proposed structures can be easily realized on an integrated photonic platform and will enable many applications that benefit from the advantages of both BICs and anti-*PT* symmetry, such as ultralow-linewidth lasing, enhanced optical sensing, and optical signal processing.

## Results

### Theoretical analysis

Figure 1 shows the schematic of our proposed structure, which consists of $N$ parallel cascaded microring resonators with their intrinsic resonant frequencies $\omega_1, \omega_2, \ldots, \omega_N$. The distances between any two adjacent resonators are equal and large, such that any resonator is coupled with the others indirectly through the bus waveguides. Assuming that the coupling rates between the resonators and bus waveguides are $\gamma_c/2$ and the intrinsic loss rates of the resonators are all $\gamma_i/2$, the effective Hamiltonian of the system is expressed as (see the Supplementary Information)

$$H = \begin{bmatrix} \Delta_1 - i\gamma_c & -i\gamma_c e^{i\theta} & \cdots & -i\gamma_c e^{(N-1)i\theta} \\ -i\gamma_c e^{i\theta} & \Delta_2 - i\gamma_c & \cdots & -i\gamma_c e^{(N-2)i\theta} \\ \vdots & \vdots & \ddots & \vdots \\ -i\gamma_c e^{(N-1)i\theta} & -i\gamma_c e^{(N-2)i\theta} & \cdots & \Delta_N - i\gamma_c \end{bmatrix}, \quad (1)$$

where $\Delta_m$ ($= \omega_m - \omega_0$) is the frequency detuning of the $m$th resonator from the center frequency of $N$ resonators with $\omega_0$ equal to $(\omega_1 + \omega_2 + \ldots + \omega_N)/N$ and $\theta$ is the phase shift between any two adjacent resonators. This system satisfies the *APT*-symmetry requirement when $\theta = n\pi$ and $\Delta_m + \Delta_{N+1-m} = 0$ with $n$ being an integer. When $\theta = 2n\pi$ and $\exp(i\theta) = 1$, the effective Hamiltonian can be simplified as:

$$H = \begin{bmatrix} \Delta_1 - i\gamma_c & -i\gamma_c & \cdots & -i\gamma_c \\ -i\gamma_c & \Delta_2 - i\gamma_c & \cdots & -i\gamma_c \\ \vdots & \vdots & \ddots & \vdots \\ -i\gamma_c & -i\gamma_c & \cdots & \Delta_N - i\gamma_c \end{bmatrix}. \quad (2)$$

Next, we analyzed the eigenfrequencies of such an *APT*-symmetric system. With different numbers ($N$) of resonators, the eigenvalues ($\sigma$) of the Hamiltonian behave differently.

Eigenfrequencies ($\omega$) of such an *APT*-symmetric system are $\omega_0 - i\gamma_i/2 + \sigma$. When $N$ (= $2k$) is an even number, the eigenvalues of the Hamiltonian satisfy

$$\left[\left(\sigma^2 - \Delta_1^2\right)\cdots\left(\sigma^2 - \Delta_k^2\right)\right]\left\{1 + 2i\gamma_c\sigma\left[1/\left(\sigma^2 - \Delta_1^2\right) + \cdots + 1/\left(\sigma^2 - \Delta_k^2\right)\right]\right\} = 0. \tag{3}$$

It shows that the eigenvalues of the system are determined by the frequency detuning of each resonator. When $\Delta_1 = \Delta_2 = \ldots = \Delta_k = \Delta$, the eigenvalues satisfy

$$(\sigma - \Delta)^{k-1}(\sigma + \Delta)^{k-1}(\sigma^2 + 2ik\gamma_c\sigma - \Delta^2) = 0. \tag{4}$$

Among all the eigenvalues, two of them satisfy $\sigma^2 + 2ik\gamma_c\sigma - \Delta^2 = 0$. Therefore, the two eigenfrequencies are $\omega_\pm = \omega_0 - i\gamma_i/2 - ik\gamma_c \pm (\Delta^2 - k^2\gamma_c^2)^{1/2}$, which behave similarly as those of *APT*-symmetric systems with two resonators[20, 21]. The spontaneous phase transition occurs at $\Delta = k\gamma_c$. The two eigenmodes take the same resonant frequency but different decay rates in the *APT*-symmetric phase ($\Delta < k\gamma_c$) while taking different resonant frequencies but the same decay rate in the *APT*-broken phase ($\Delta > k\gamma_c$). System performance can also be enhanced based on the system with a higher number of resonators. For example, non-Hermitian systems can be used for high-sensitivity sensing. The sensitivity of such system working at exceptional points is defined as the change in eigenfrequencies when the frequency detuning $\Delta$ deviates from $k\gamma_c$ by $\varepsilon$ ($\varepsilon \ll k\gamma_c$). Increasing resonator numbers can also enhance the sensitivity, which is approximately proportional to $(k\gamma_c\varepsilon)^{1/2}$. The other eigenfrequencies ($\omega = \omega_0 - i\gamma_i/2 \pm \Delta$) belong to the Fabry–Pérot (FP) BICs. However, when $N$ (= $2k + 1$) is an odd number, the eigenvalues of the Hamiltonian satisfy

$$1 + 2i\gamma_c\sigma\left[1/\left(\sigma^2 - \Delta_1^2\right) + 1/\left(\sigma^2 - \Delta_2^2\right) + \cdots + 1/\left(\sigma^2 - \Delta_k^2\right) + 1/2\sigma^2\right] = 0. \tag{5}$$

Similar to the discussion above, when $\Delta_1 = \Delta_2 = \ldots = \Delta_k = \Delta$, Eq. (5) reduces to

$$\sigma^3 + (2k+1)i\gamma_c\sigma^2 - \Delta^2\sigma - i\gamma_c\Delta^2 = 0, \tag{6}$$

where exceptional points do not exist.

We also analyzed the FP BICs in the *APT*-symmetric system in detail. When $N = 2$, FP BICs can be realized if the intrinsic frequencies of resonators are equal and the phase shift is integer multiple of $\pi$[40]. The phase shift naturally satisfies the FP resonance requirement since *APT*-symmetric systems require that $\theta = 2n\pi$. When $N > 2$, we can also obtain cascaded FP BICs if there is no resonance detuning ($\Delta = 0$). The eigenfrequencies of the resonator system thus become $\omega_0 - i\gamma_i/2$ and $\omega_0 - i\gamma_i/2 - iN\gamma_c$. These eigenmodes with frequencies $\omega_{BIC}$ (= $\omega_0 - i\gamma_i/2$) are FP BICs and the degeneracy of $\omega_{BIC}$ is $N − 1$. For the FP BIC, the light is trapped between the $N$ resonators, where the resonant radiation into the continuum through the radiation channels (bus waveguides) interferes destructively with each other. In this case, the quality factor of this cascaded FP BIC ($\omega_0/\gamma_i$) is relatively high, and the lifetime becomes infinitely long when the intrinsic loss rate $\gamma_i/2$ is negligible.

We further analyzed the transmission characteristics for the parallel cascaded-resonator system based on the temporal coupled-mode theory (see the Supplementary Information):

$$t = \left\{ 1 - \gamma_c \left[ 1/(i\delta\omega_1 - \gamma_i/2) + 1/(i\delta\omega_2 - \gamma_i/2) + \cdots + 1/(i\delta\omega_N - \gamma_i/2) \right] \right\}^{-1}, \tag{7}$$

where $T (= |t|^2)$ is the transmission rate and $\delta\omega_j (= \omega_p − \omega_j)$ is the frequency detuning between the probe frequency $\omega_p$ and the resonant frequency of the $j$th resonator ($\omega_j$). If $\gamma_c \gg \delta\omega_j \gg \gamma_i$, we can find that $T$ approaches 0 when the probe frequency coincides with any one of the intrinsic resonant frequencies. There also exists a peak ($T$ approaching 1) between any two different resonant frequencies in the transmission spectra. Figure 2 shows the numerically calculated transmission spectra for a system consisting of 4 cascaded resonators. When the intrinsic resonant frequencies of the 4 resonators are different, there are 4 dips at the resonant frequencies and 3 peaks sandwiched by the resonant dips. When two resonant frequencies are equal ($\omega_2 = \omega_3$), the second and third dips merge. When $\omega_1 = \omega_2 = \omega_3 = \omega_4$ the cascaded FP BIC exists, and the transmission rate is 0 at the resonant frequency. Such transmission spectra are similar to those of electromagnetically induced transparency (EIT)[40]. Our results also prove that the artificial transparency windows, which are the peaks in the transmission spectra, can be dynamically controlled by finely tuning the intrinsic resonant frequencies of the resonators. The quality factor and position of transmission peaks can also be tuned by changing the intrinsic resonant frequencies of resonators.

In case that the cascaded resonators are close to each other causing direct coupling ($\kappa$) between them, the effective Hamiltonian for an $N = 2$ system can be expressed as

$$H = \begin{bmatrix} 0 & \kappa \\ \kappa & 0 \end{bmatrix} + \begin{bmatrix} \Delta - i\gamma_c & -i\gamma_c \\ -i\gamma_c & -\Delta - i\gamma_c \end{bmatrix}, \tag{8}$$

which has the eigenvalues $\sigma_\pm = -i\gamma_c \pm [\Delta^2 + (\kappa - i\gamma_c)^2]^{1/2}$. When $\Delta = 0$, the eigenfrequencies are $\omega_0 - i\gamma_i/2 - \kappa$ and $\omega_0 - i\gamma_i/2 + \kappa - 2i\gamma_c$. The FP BIC still exists with $\omega_{BIC} = \omega_0 - i\gamma_i/2 - \kappa$. The two eigenmodes no longer have the same resonant frequency because the direct coupling breaks the *APT* symmetry. The eigenfrequencies are not degenerate at the exceptional point ($\Delta = \gamma_c$) but are separated with $\Delta\omega = 2(\kappa^2 - 2i\kappa\gamma_c)^{1/2}$. The transmission spectra have also changed significantly: (see the Supplementary Information)

$$t = e^{i\theta} \frac{(i\delta\omega_1 - \gamma_i/2)(i\delta\omega_2 - \gamma_i/2) + 2\kappa\gamma_c \sin\theta + \kappa^2}{(i\delta\omega_1 - \gamma_i/2)(i\delta\omega_2 - \gamma_i/2) - \gamma_c[i(\delta\omega_1 + \delta\omega_2) - \gamma_i] - 2i\gamma_c e^{i\theta}(\kappa + \gamma_c \sin\theta) + \kappa^2}. \quad (9)$$

Figure 3 plots the transmission spectra with different phase shifts $\theta$ and direct coupling rates $\kappa$. For a system consisting of two identical resonators, FP BICs exist when $\theta = 2n\pi$, giving rise to only one dip in the transmission spectra [Fig. 3(a)]. The resonant frequency of the FP BIC shifts by $\kappa$ when the direct coupling $\kappa$ is introduced. However, for $\theta = 2n\pi + \pi/2$, the dip splits into two separate dips as $\kappa$ increases [Fig. 3(b)]. When $\theta = 2n\pi$ and the resonant frequencies of the two resonators are different, an optical analog of the EIT phenomenon appears with $\kappa = 0$. The two dips are further separated with the increase of $\kappa$, evolving into a high-$Q$ dip and a low-$Q$ dip [Fig. 3(c)]. For $\theta = 2n\pi + \pi/2$, the two dips are also further separated as $\kappa$ increases but with the same linewidth [Fig. 3(d)]. These phenomena as shown in the transmission spectra in Fig. 3 enable versatile and flexible applications in optical switching and filtering. With such a parallel cascaded-resonator system, we can control the transmission (or the "on/off" state in optical switching) at a certain wavelength by tuning any one of these parameters: intrinsic resonant frequencies, phase shift $\theta$, and coupling rates $\kappa$.

**Numerical simulation**

Based on the theoretical calculation, we adopted a two-dimensional finite-element method in COMSOL to simulate the *APT*-symmetric optical system. Generalizations of our two-dimensional simulation to three-dimensional structures are straightforward. Without loss of generality, we designed the structures based on a silicon integrated photonic platform. All the microring resonators have a radius of 3.1 μm and a waveguide width of 0.4 μm. The width of the bus waveguides is 0.4 μm and the gaps between the waveguide and the ring resonators are 0.1 μm. Silicon has a negligible intrinsic absorption rate ($\gamma_i = 0$) in the communication band (~1550 nm). The resonant frequency is tuned by slightly changing the microring's inner radius. We analyzed the FP BICs and the evolution of the eigenfrequencies with $N$ (= 2, 3, 4) cascaded resonators.

Figures 4(a) and 4(b) plot the change in eigenfrequencies of the *APT*-symmetric system with 2 resonators. The requirement of FP BIC is satisfied when $\Delta = 0$ ($\omega_1 = \omega_2$). In this case, the light is trapped between the two resonators, as shown in the insets of Fig. 4(b). The exceptional points can be observed from the evolution of the eigenfrequencies when $\Delta \neq 0$. The system works in the *APT*-symmetric phase and the two eigenmodes share the same resonant frequency but have different decay rates when $\Delta < \gamma_c$. The system works in the *APT*-broken phase and the two eigenmodes share the same coupling loss rates but have asymmetric optical field distributions when $\Delta > \gamma_c$. The two eigenmodes coalesce at the exceptional point ($\Delta = \gamma_c$) where their eigenfrequencies are degenerate.

The cases are different for systems with a larger number of resonators. Figures 4(c) and 4(d) show the simulated results for an *APT*-symmetric system with $N = 3$. Cascaded FP BIC also exists when $\Delta = 0$ ($\omega_1 = \omega_2 = \omega_3$). There are two high-$Q$ modes with asymmetric field distributions and one low-$Q$ mode when $\Delta \neq 0$. The evolution behavior of the eigenfrequencies indicates that the eigenmodes of the parallel cascaded-resonator system with an odd $N$ do not merge and no exceptional point exists, which agrees with the analytical results. Figures 4(e) and 4(f) show the simulated results for an *APT*-symmetric system with $N = 4$. When all the resonators share the same resonant frequency, the cascaded FP BIC is formed where the light is confined to the middle two resonators, as shown in an inset of Fig. 4(f). Similar to the case of $N = 2$, the two eigenmodes take the same resonant frequency but different decay rates in the *APT*-symmetric phase ($\Delta_1 = \Delta_2 < k\gamma_c$) while taking the same decay rate and exhibiting asymmetric field distributions in the *APT*-broken phase ($\Delta_1 = \Delta_2 > k\gamma_c$). The other two eigenmodes with purely real eigenfrequencies are the FP BICs formed between the first and last resonators with $\Delta_1 = \Delta_2$. (see the Supplementary Information for the simulation of systems with more than 4 resonators)

**Effects of experimental imperfections**

The proposed structure can be realized on an integrated photonic platform. In real systems, the phase shifts between the resonators and the intrinsic resonant frequencies can be tuned precisely by the electro-optic or thermo-optic effect. However, experimental imperfections are unavoidable. Here, we investigated the effects of imperfections in our cascaded-resonator systems. For a system consisting of two resonators where the phase shift $\theta$ is not equal to $n\pi$, the coupling rates between the bus waveguides and the two resonators are different. The effective Hamiltonian of this system becomes

$$H = \begin{bmatrix} \Delta - i\gamma_{c1} & -i\sqrt{\gamma_{c1}\gamma_{c2}}\,e^{i\theta} \\ -i\sqrt{\gamma_{c1}\gamma_{c2}}\,e^{i\theta} & -\Delta - i\gamma_{c2} \end{bmatrix}. \tag{10}$$

The eigenfrequencies of such a system are $\omega_{\pm} = \omega_0 - i\gamma_i/2 - i(\gamma_{c1} + \gamma_{c2})/2 \pm [-(\gamma_{c1} - \gamma_{c2})^2/4 + \Delta^2 - i\Delta(\gamma_{c1} - \gamma_{c2}) - \gamma_{c1}\gamma_{c2}\exp(i\theta)]^{1/2}$.

Based on the Hamiltonian, we first considered the fabrication imperfections affecting the gaps between the bus waveguides and the two resonators rendering $\gamma_{c1} \ne \gamma_{c2}$. For $\theta = 2n\pi$, the eigenfrequencies become $\omega_{\pm} = \omega_0 - i\gamma_i/2 - i(\gamma_{c1} + \gamma_{c2})/2 \pm [-(\gamma_{c1} + \gamma_{c2})^2/4 - i\Delta(\gamma_{c1} - \gamma_{c2})]^2$. The FP BIC ($\omega_{BIC} = \omega_0 - i\gamma_i/2$) still exists when $\Delta = 0$. The FP BIC also exists for a system consisting of $N$ resonators with $\omega_{BIC} = \omega_0 - i\gamma_i/2$, which indicates that the FP BIC exists irrespective of the fabrication imperfections about the gaps between the bus waveguides and the resonators. However, the eigenfrequencies are not degenerate at the exceptional point in the presence of imperfections when $\Delta \ne 0$. $\Delta\omega$ ($\omega_+ - \omega_-$) at the exceptional point, which vanishes in the ideal case, increases as the degree of imperfections ($\gamma_{c1} - \gamma_{c2}$) increases.

We also analyzed the effect of $\theta$ deviation from $n\pi$. When $\gamma_{c1} = \gamma_{c2}$ and $\Delta = 0$, the eigenfrequencies become $\omega_{\pm} = \omega_0 - i\gamma_i/2 - i\gamma_c[1 \pm \exp(i\theta)]$. Deviating from the FP BIC, the eigenmode transits to a state near BIC with the eigenfrequency $\omega_{NBIC}$ equal to $\omega_0 - i\gamma_i/2 - i\gamma_c[1 - \exp(i\theta)]$. In this case, the FP resonance condition is broken. Figure 5(a) plots the numerically calculated deviation of the imaginary part of eigenfrequencies as a function of $\theta$ and $N$ for systems consisting of $N$ resonators. It indicates that the quality factors of the eigenmodes in the system consisting of more resonators have a higher tolerance to the $\theta$ deviation. For $\Delta \ne 0$, the evolution of the eigenfrequencies also deviates from that of the $APT$-symmetric system. The eigenfrequencies can be approximated as $\omega_{\pm} = \omega_0 - i\gamma_i/2 - i\gamma_c \pm \gamma_c(i-1)\theta^{1/2}$ when $\theta - n\pi \ll 1$ at the exceptional point ($\Delta = \gamma_c$). $\Delta\omega$ ($= \omega_+ - \omega_-$) at the exceptional points, which consists of both real and imaginary parts, increases as the $\theta$ deviation increases. Figure 5(b) plots the effect of $\theta$ deviation on imaginary parts of eigenfrequencies. Figure 5(c) plots the calculated real part of $\Delta\omega/2\pi\gamma_c$ as a function of $\theta$ and $N$. It indicates that the evolution of the eigenfrequencies of a system consisting of more resonators deviates more from its ideal case with the same $\theta$ deviation.

**Discussion**

In summary, we have theoretically shown that Fabry–Pérot BICs and anti-$PT$ symmetry can be realized simultaneously in a parallel cascaded-resonator system with interesting features. First, the characteristic transmission of such systems, such as the optical analog of electromagnetically induced transparency, enables versatile applications in optical filtering and switching. Second, the sensitivity at the exceptional points is enhanced in a system with more cascaded resonators. Last, taking potential imperfections into consideration, the Fabry–Pérot BICs can be realized regardless of the fabrication imperfections causing variations of the gaps between the bus waveguides and the resonators, and a system consisting of more resonators is more tolerant to the deviation of the phase shift between resonators. Our proposed structures that simultaneously achieve BICs and anti-*PT* symmetry on an integrated photonic platform will lead to new applications that benefit from the advantages of both BICs and anti-*PT* symmetry.

**Acknowledgements**

This work was supported by Research Grants Council of Hong Kong (14208717, 14206318, 14208421).

**Author contributions**

X.S. conceived the idea. J.Z. performed theoretical derivation, numerical simulation, and data analysis with the help of Z.F.. All authors worked together to write the manuscript. X.S. supervised the project.

**Competing interests**

The authors declare no competing interests.

**Data availability**

The data that support the findings of this study are available from the corresponding author upon reasonable request.

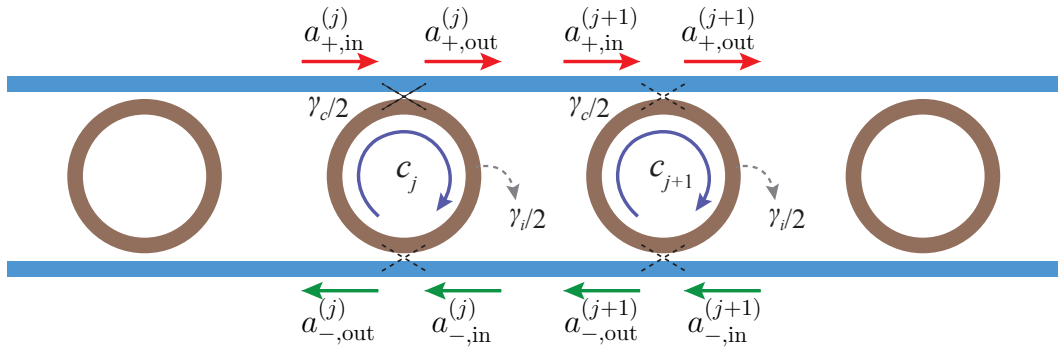

**Fig. 1 | Schematic of a parallel cascaded-resonator system for realizing both *APT* symmetry and Fabry–Pérot BICs.**

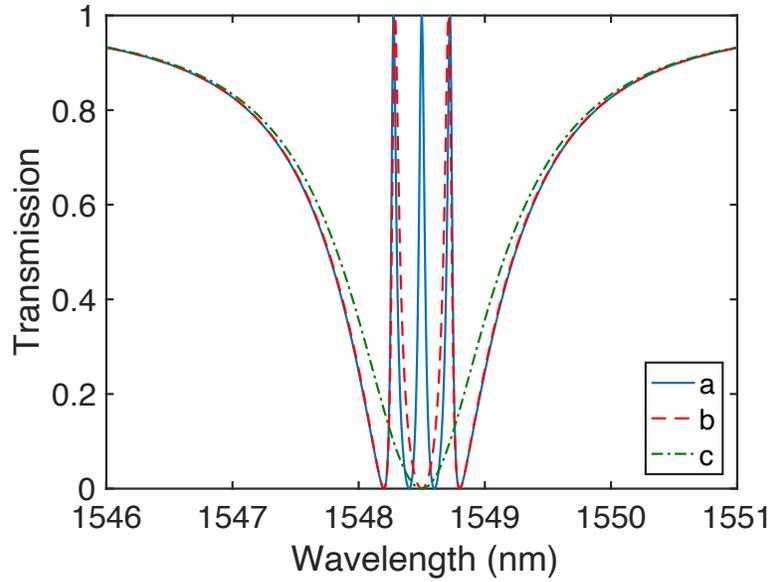

**Fig. 2 | Theoretically calculated transmission spectra for systems consisting of 4 cascaded resonators.** The resonant wavelengths are set to be (a) $\lambda_1$ = 1548.2 nm, $\lambda_2$ = 1548.4 nm, $\lambda_3$ = 1548.6 nm, $\lambda_4$ = 1548.8 nm, (b) $\lambda_1$ = 1548.2 nm, $\lambda_2$ = $\lambda_3$ = 1548.5 nm, $\lambda_4$ = 1548.8 nm, and (c) $\lambda_1$ = $\lambda_2$ = $\lambda_3$ = $\lambda_4$ = 1548.5 nm. The distances between any two adjacent resonators are equal and large enough.

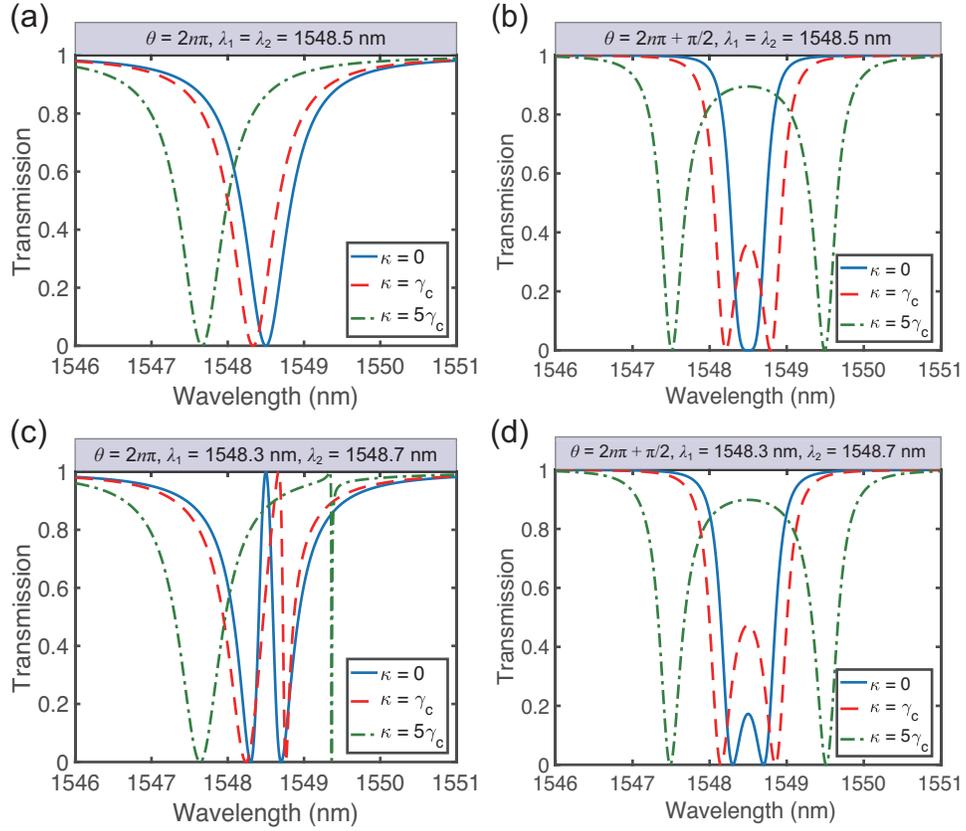

**Fig. 3 | Theoretically calculated transmission spectra for a system consisting of two resonators that are close to each other.** (**a**) $\theta = 2n\pi$, $\lambda_1 = \lambda_2 = 1548.5$ nm. (**b**) $\theta = 2n\pi + \pi/2$, $\lambda_1 = \lambda_2 = 1548.5$ nm. (**c**) $\theta = 2n\pi$, $\lambda_1 = 1548.3$ nm, $\lambda_2 = 1548.7$ nm. (**d**) $\theta = 2n\pi + \pi/2$, $\lambda_1 = 1548.3$ nm, $\lambda_2 = 1548.7$ nm.

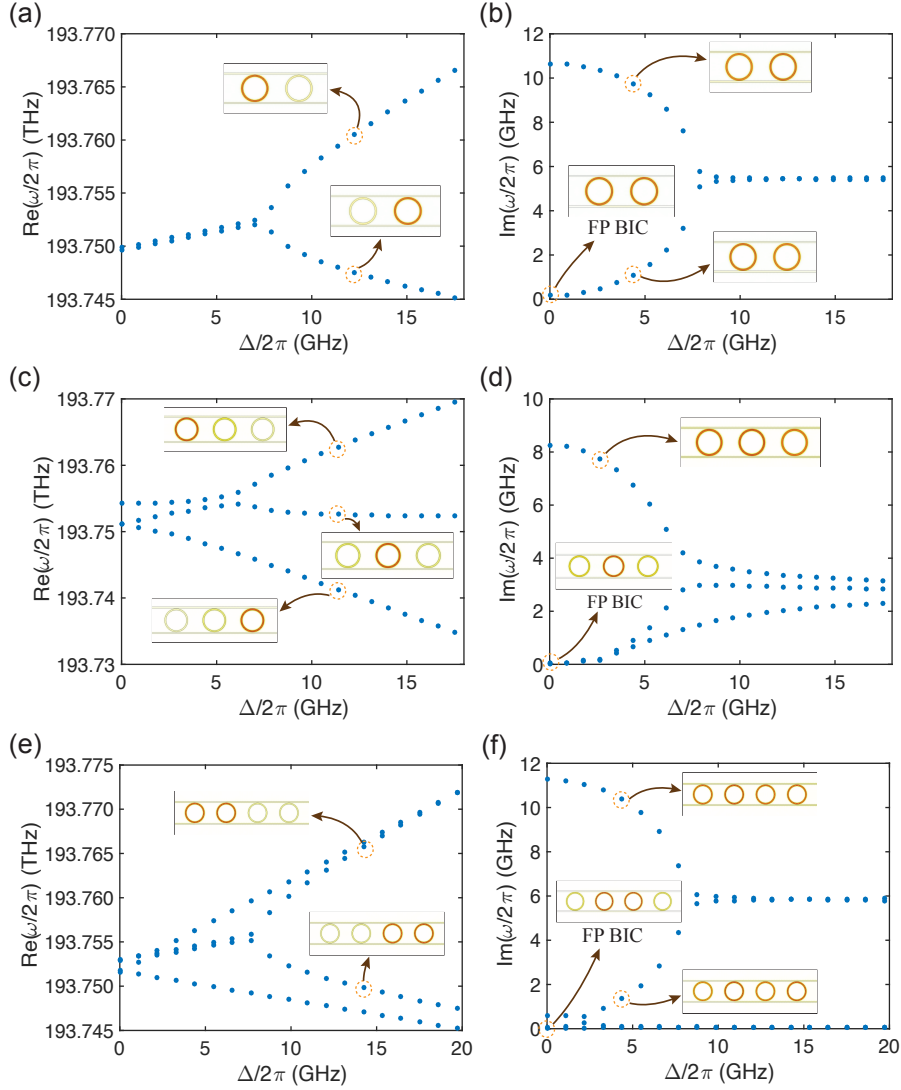

**Fig. 4 | Simulated evolution of the eigenfrequencies for the parallel cascaded-resonator system.** (**a**),(**b**) Real and imaginary part of the eigenfrequencies for the $N = 2$ system. The insets in (**a**) show the field distributions of the two eigenmodes in the *APT*-broken phase ($\Delta = 1.6471\gamma_c$), while the insets in (**b**) show those in the *APT*-symmetric phase ($\Delta = 0.5882\gamma_c$) as well as that of the FP BIC. (**c**),(**d**) Real and imaginary part of the eigenfrequencies for the $N = 3$ system. The insets in (**c**) show the field distributions of the 3 eigenmodes when $\Delta = 1.5294\gamma_c$, while the insets in (**d**) show that of a low-*Q* state ($\Delta = 0.3529\gamma_c$) as well as that of the FP BIC. There is no exceptional point in this case. (**e**),(**f**) Real and imaginary part of the eigenfrequencies for the $N = 4$ system. The insets in (**e**) show the field distributions of two eigenmodes in the *APT*-broken phase ($\Delta = 1.625k\gamma_c$), while the insets in (**f**) show those in the *APT*-symmetric phase ($\Delta = 0.5k\gamma_c$) as well as that of the cascaded FP BIC.

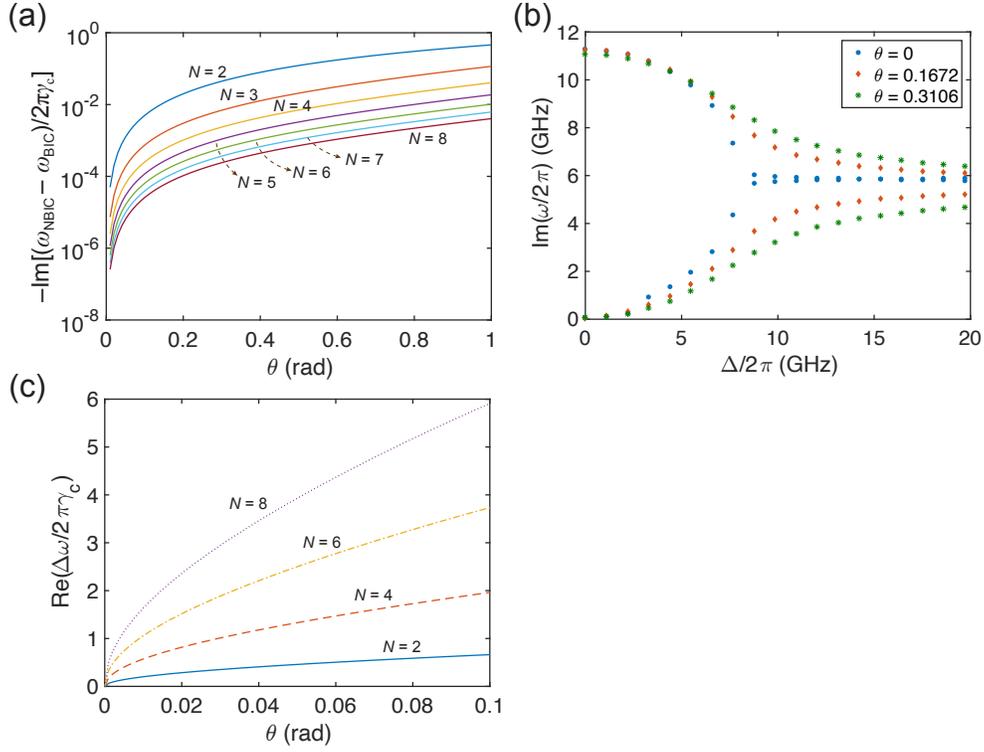

**Fig. 5 | Numerical results of potential experimental imperfections.** (**a**) Numerically simulated deviation Im[$(\omega_{\text{NBIC}} - \omega_{\text{BIC}})/2\pi\gamma_c$] for systems consisting of $N$ = 2, 3, …, 8 resonators. (**b**) Numerically simulated effect of $\theta$'s deviation on the *APT* symmetry for a system consisting of 4 resonators. (**c**) Numerically simulated deviation Re($\Delta\omega/2\pi\gamma_c$) at exceptional points for systems consisting of $N$ = 2, 4, 6, 8 resonators.

17